\documentclass[UTF8,a4paper]{article}
\usepackage{amsmath}
\usepackage{graphicx}
\usepackage{subfigure}
\usepackage{epstopdf}
\usepackage{geometry}
\usepackage{ulem}
\usepackage{indentfirst}
\usepackage{amsfonts,amssymb,dsfont}
\usepackage{setspace}
\usepackage{threeparttable}
\usepackage{amsmath,mathtools,amsthm}
\usepackage{array}
\usepackage{extarrows}
\usepackage{upgreek}

\makeatletter
\renewcommand*\env@matrix[1][\arraystretch]{%
  \edef\arraystretch{#1}%
  \hskip -\arraycolsep
  \let\@ifnextchar\new@ifnextchar
  \array{*\c@MaxMatrixCols c}}
\makeatother
\begin{document}


\title{\bf Integer quantum Hall conductivity and longitudinal conductivity in silicene under the electric field and magnetic field}
\author{Chen-Huan Wu
\thanks{chenhuanwu1@gmail.com}
\\Key Laboratory of Atomic $\&$ Molecular Physics and Functional Materials of Gansu Province,
\\College of Physics and Electronic Engineering, Northwest Normal University, Lanzhou 730070, China}

\maketitle
\vspace{-30pt}
\begin{abstract}
\begin{large}
We investigate the integer Hall conductivity and longitudinal conductivity of silicene under the magnetic field, electric field, and 
exchange field in this letter.
We focus not only on the low-temperature and $\delta$-function impurities (i.e., independent of the scattering momentum) case,
which only exist the intra-Landau level transition,
but also on the case of inter-Landau level transition which also with the non-elastic scattering.
The resulting longitudinal conductivity is very different with the intra-Landau level one at low-temperature.
The exprssions of the Hall conductivity, longitudinal conductivity, valley contributed Hall conductivity, and the spin or valley Hall conductivity,
are deduced in this letter.
We also compute the dynamical polarization under the magnetic field 
which is a important quantity and has many exciting and novel properties, 
and with the screened scattering due to the charged impurities.
The polarization function here is also related to the Landau level index under the magnetic field,
and shows step-like feature rather than the logarithmically divergenas which appears for the zero-magnetic field case,
and it's also naturally related to the conductivity in silicene.
The generalized Laguerre polynomial is used in both the longitudinal conductivity and dynamical polarization function
under magnetic field.

\end{large}
\begin{large}

\end{large}
\end{abstract}
\begin{large}
\section{Introduction}

Silicene, 
the silicon version of the graphene,
has attachted much attentions both experimentally
and theoretically since it's successfully synthesized
together with it's bilayer form or nanoribbon form\cite{Feng B},
and it has the properties of both the topological insulator (TI) and semimetal,
which provides possibility for the abundant phase transitions\cite{Ezawa M}.
The low-energy dynamics of silicene can be well described by the Dirac-theory.
The silicene is also a $3p$-orbital-based materials with the noncoplanar low-buckled (with a buckle about $0.46$ \AA\ due to the hybridization between the 
$sp^{2}$-binding 
and the $sp^{3}$-binding (which the bond angle is $109.47^\text{o}$) and 
that can be verified by thr Raman spectrum 
which with the intense peak at 578 cm$^{-1}$ larger than the planar one and the $sp^{3}$-binding one
\cite{Tao L},
and thus approximately forms two surface-effect (like the thin ferromagnet matter) lattice structure.
The bulked structure not only breaks the lattice inversion symmetry,
but also induce a exchange splitting between the upper atoms plane and the lower atom plane and thus forms a emission geometry which allows the
optical interband transitions, which for the graphene can happen only upon a FM substrate.
The FM or AFM order can be formed in monolayer silicene by the magnetic proximity effect
that applying both the perpendicular electric field and in-plane FM or AFM field.
Silicene also has much stronger intrinsic spin-orbit coupling (SOC) and stronger interlayer interaction compared to the graphene due to its
heavier atom mass and low-bulked structure, respectively.
That also plays a important role during the phase transitions of silicene.
In this letter, we mainly investigate the integer Hall conductivity and longitudinal conductivity of silicene under the magnetic field, electric field, and 
exchange field.
The results reveal the Dirac-mass and magnetic field dependence of the conductivities in the presence of the impurities,
which behave like a $\delta$-function in the long-wavelength limit (${\bf q}\rightarrow 0$).

\section{Theory model and method}

In tight-binding model,
the Hamiltonian of monolayer silicene in low-energy Dirac theory is
\begin{equation} 
\begin{aligned}
H=&t\sum_{\langle i,j\rangle ;\sigma}c^{\dag}_{i\sigma}c_{j\sigma}
+i\frac{\lambda_{{\rm SOC}}}{3\sqrt{3}}\sum_{\langle\langle i,j\rangle\rangle ;\sigma\sigma'}\upsilon_{ij}c^{\dag}_{i\sigma}\sigma^{z}_{\sigma\sigma'}c_{j\sigma'}
- i\frac{2R}{3}\sum_{\langle\langle i,j\rangle\rangle ;\sigma\sigma'}c^{\dag}_{i\sigma}(\mu \Delta({\bf k}_{ij})\times {\bf e}_{z})_{\sigma\sigma'}c_{i\sigma'}\\
&+iR_{2}(E_{\perp})\sum_{\langle i,j\rangle;\sigma\sigma'}c^{\dag}_{i\sigma}(\Delta({\bf k}_{ij})\times {\bf e}_{z})_{\sigma\sigma'}c_{i\sigma'}
-\frac{\overline{\Delta}}{2}\sum_{i\sigma} c^{\dag}_{i\sigma}\mu E_{\perp}c_{i\sigma}\\
&+M_{s}\sum_{i\sigma}c^{\dag}_{i\sigma}\sigma_{z}c_{i\sigma}
+M_{c}\sum_{i\sigma}c^{\dag}_{i\sigma}c_{i\sigma}
+U\sum_{i}\mu n_{i\uparrow}n_{i\downarrow},
\end{aligned}
\end{equation}
where $t=1.6$ eV is the nearest-neoghbor hopping which contains the contributions from both the $\pi$ band 
and $\sigma$ band.
The gap function is $\Delta({\bf k})={\bf d}({\bf k})\cdot{\pmb \sigma}$ which in a coordinate independent but spin-dependent representation.
The ${\bf k}$-dependent unit vector ${\bf d}({\bf k})$ here has
${\bf d}({\bf k})=[t'_{SOC}{\rm sin}k_{x},t'_{SOC}{\rm sin}k_{y},M_{z}-2B(2-{\rm cos}k_{x}+{\rm cos}k_{y})]$
for the BHZ model, where $B$ is the BHZ model
-dependent parameter and $M_{z}$ the Zeeman field term which dominate the surface magnetization
but can be ignore when a strong electric field or magnetic field is applied.
$\langle i,j\rangle$ and $\langle\langle i,j\rangle\rangle $ denote the nearest-neighbor (NN) pairs and the next-nearest-neighbor (NNN) pairs, respectively. 
$\mu=\pm 1$ denote the $A$ ($B$) sublattices. 
Here ${\bf d}({\bf k}_{ij})=\frac{{\bf d}_{ij}}{|{\bf d}{ij}|}$ is the NNN hopping vector.
$\lambda_{SOC}=3.9$ meV is the intrinsic spin-orbit coupling (SOC) strength which is much larger than the monolayer graphene's (0.0065 meV\cite{Guinea F}).
$R$ is the small instrinct Rashba-coupling due to the low-buckled structure, which is related to the helical bands
(helical edge states) and the SDW in silicene, and it's disappear in the Dirac-point ($k_{x}=k_{y}=0$).
$R_{2}(E_{\perp})$ is the extrinsic Rashba-coupling induced by the electric field.
The existence of $R$ breaks U(1) spin conservation (thus the $s^{z}$ is no more conserved) and the mirror symmetry of silicene lattice.
The exchange field $M$ is applied perpendicular to the silicene
and induce the staggered potential term, and it can be rised by proximity coupling to the ferromagnet.
$\upsilon_{ij}=({\bf d}_{i}\times{\bf d}_{j})/|{\bf d}_{i}\times{\bf d}_{j}|=1(-1)$ 
when the next-nearest-neighboring hopping of electron is toward left (right),
with ${\bf d}_{i}\times{\bf d}_{j}=\sqrt{3}/2(-\sqrt{3}/2)$.
The last term is the Hubbard term with on-site interaction $U$ which doesn't affects the bulk gap here but affects the edge gap.

The low-temperture longitudunal in-plane conductivity (diagonal) of silicene with the dominating elastic scattering due to the charged impurity
in linear response theory is\cite{Charbonneau M}
\begin{equation}
\begin{aligned}
\sigma_{xx}=\sigma_{yy}=\frac{\beta e^{2}}{S}\sum_{m}f_{m}(1-f_{m})\frac{\langle m|v_{x}|m\rangle\langle m|v_{y}|m\rangle}{\omega+i\delta+i\Gamma}
\end{aligned}
\end{equation}
where $S=3\sqrt{3}a^{2}/2$ is the area of unit cell (Wigner-Seitz cell), 
$\beta$ is the inversed temperature, $m$ denotes the electron state,
$\omega=(2n+1)\pi/\beta$ is the fermionic Matsubara frequency where $\beta$ is the inverse temperature.
$v_{x}=\frac{\partial}{\hbar\partial k_{x}}$ is the velocity operator.
Here we use the retarded form by the analytical continuation as $i\omega\rightarrow \omega+i\delta$.
The longitudunal conductivity is related to the interband transmission, and the carriers collisions (especially under the magnetic field),
and it also related to the screened Coulomb scattering by the charged impurity with the transfered cyclotron orbit when under the magnetic field
with
the cyclotron resonance frequency $\omega_{c}=\frac{\sqrt{2} v_{F}}{\ell_{B}}$,
where $\ell_{B}=\sqrt{\hbar c/|eB|}$ is the magnetic length which plays the role of quantized cyclotron orbit radius
in lowest Landau level (LLL) (n=0) $R_{0}=\ell_{B}$ and the quantized cyclotron orbit radius for $n\neq 0$ is $R_{n}=\sqrt{2n}\ell_{B}$.
In this case, the kinetic energy of a single-electron is $\sim \hbar\omega_{c}$\cite{Kotov V N}.
It's also found that, with the increase of chemical potential,
the spectral weight of intraband transition is rised for the real part of longitudunal in-plane conductivity $\sigma_{xx}$\cite{Wu C H,Tabert C J}
The scattering rate $\Gamma$ here is defined as
\begin{equation} 
\begin{aligned}
\Gamma=\frac{1}{2\tau}=\frac{\pi n}{\hbar}V^{2},
\end{aligned}
\end{equation}
where $\tau$ is the quasiparticle lifetime.
Here the charged impurity density $n$ is momentum-independent for the single-impurity case.
If the SOC is taken into consider in the collision process of the impurity scattering as done in the explores of spin-Hall effect\cite{Dyakonov M I}
with the presence of electric currence and the spin currence,
the $\Gamma$ becomes $<\frac{1}{2\tau}$ and thus the DOS may be increased.
In the domination of impurity scattering and with a certain impurity concentration, the 
spin currence can be described by angle $\theta$ in Maxwell theory.
with the spin-palarized currence perpendicular to the applied electric field which exhibit the quantum quantum anomalous Hall (QAH) effect\cite{Wu C H2}.

For the magneto conductivity, which is contributed by the diagonal (longitudinal) and nondiagonal (transverse) Hall conductivity.
The former one is related to the interband transition and the localized state Shubnikov de Hass oscillations,
and it may leads to the transfer of cyclotran orbit which with cyclotron resonance frequency $\omega_{c}$
by the scattering in the presence of charged impurities.
The magnetic field ${\bf B}=(B_{x},0,B_{z})=\nabla\times{\bf A}$ is applied perpendicular to the silicene and 
with the Landau gauge ${\bf A}=(-B_{z}y,0,B_{x}y)$,
and the momentum can be replaced by the covariant momentum ${\bf P}=\hbar{\bf P}+\frac{e}{c}{\bf A}$,
which ${\bf P}=\hbar(\frac{y}{\ell_{B}}-\ell_{B}k_{x}+\partial_{y})$ and ${\bf P}^{\dag}=\hbar(\frac{y}{\ell_{B}}-\ell_{B}k_{x}-\partial_{y})$.

For the transverse in-plane conductivity (non-diagonal) under the magnetic field, it's
\begin{equation}
\begin{aligned}
\sigma_{xy}=\frac{i\hbar e^{2}}{\omega_{c}S_{B}}\sum_{E_{n'}>E_{n},s_{z},\eta}[f(E_{n})-f(E_{n'})]\frac{\langle n|v_{x}|n'\rangle\langle n'|v_{y}|n\rangle}
{(E_{n}-E_{n'})(E_{n}-E_{n'}+\omega+i\delta+i\Gamma)},
\end{aligned}
\end{equation}
where the normalized velocity matrix elememts are
\begin{equation}
\begin{aligned}
\langle n|v_{x}|n'\rangle=\frac{v_{F}}{2}       (s(1+\frac{m_{D}^{++}}{E_{n}})^{1/2}(1-\frac{m_{D}^{+-}}{E_{n'}})^{1/2})\delta_{s_{z},s_{z}'}\delta_{t_{z},t_{z}'}(\delta_{n',n-1}+\delta_{n',n+1}),\\
\langle n'|v_{x}|n\rangle=\frac{i\eta v_{F}}{2}(-s(1-\frac{m_{D}^{+-}}{E_{n'}})^{1/2}(1-\frac{m_{D}^{++}}{E_{n}})^{1/2})\delta_{s_{z},s_{z}'}\delta_{t_{z},t_{z}'}(\delta_{n',n-1}-\delta_{n',n+1}),\\
\end{aligned}
\end{equation}
where $s,s'$ are the band index ($ss'=1$ for the intraband case and $ss'=-1$ for the interband case),
and with the Dirac-mass as
\begin{equation} 
\begin{aligned}
m_{D}^{\eta s_{z}}=\eta\lambda_{{\rm SOC}}s_{z}t_{z}-\frac{\overline{\Delta}}{2}E_{\perp}t_{z}+Ms_{z}.
\end{aligned}
\end{equation}
where $s_{z}=\pm 1$, $\eta=\pm 1$, $t_{z}=\pm 1$ denote the spin, valley, and sublattice degrees of freedom,
and in our calculations, we set $n'=n+1$, $n\ge 0$ and $n\in N$.
The Dirac-mass here, for simplicity, we assume the sublattice index $t_{z}$ change sign when the sign of valley index $\eta$ change,
while the spin index $s_{z}$ is independent of the valley transitions, 
which is since the spin flip or not during the transition is all possible and depending on the properties of the edge states and the
magnetic impurities.
Thus the superscript of Dirac-mass can be simplified as only contains $\eta$ and $t_{z}$.
The exchange field in this paper is always setted as $M=0.039$ eV.
The specific area here consider the effect of magnetic field is $S_{B}=2r_{B}$ with $r_{B}=\ell_{B}^{2}{\bf q}^{2}/2$
under Cartesian coordinate and 
has $\int{\bf q}d{\bf q}=(1/\ell_{B}^{2})\int d r_{B}$.
The eigenstates in the above velocity matrix elememt for the $n\neq 0$ Landau levels are
\begin{equation} 
\begin{aligned}
|n\rangle^{K}_{s,s_{z}}=\frac{e^{ik_{x}x}}{\sqrt{2}}
\begin{pmatrix}
(1+\frac{m_{D}^{+s_{z}}}{E_{n}})^{1/2}\Phi_{n}\\
-s(1-\frac{m_{D}^{+s_{z}}}{E_{n}})^{1/2}\Phi_{n-1}
\end{pmatrix}
\end{aligned}
\end{equation}
for the $K$ valley, and 
\begin{equation} 
\begin{aligned}
|n\rangle^{K'}_{s,s_{z}}=\frac{e^{ik_{x}x}}{\sqrt{2}}\begin{pmatrix}(1+\frac{m_{D}^{-s_{z}}}{E_{n}})^{1/2}\Phi_{n-1}\\
-s(1-\frac{m_{D}^{-s_{z}}}{E_{n}})^{1/2}\Phi_{n}
\end{pmatrix}
\end{aligned}
\end{equation}
for the $K'$ valley.
While for the zeroth Landau level, it need to be treated separately due to its specificity. They are
\begin{equation} 
\begin{aligned}
|0\rangle^{K}_{s,s_{z}}=e^{ik_{x}x}\begin{pmatrix}\Phi_{0}\\
0
\end{pmatrix}
\end{aligned}
\end{equation}
for the $K$ valley, and 
\begin{equation} 
\begin{aligned}
|0\rangle^{K'}_{s,s_{z}}=e^{ik_{x}x}\begin{pmatrix}0\\
-s\Phi_{0}
\end{pmatrix}
\end{aligned}
\end{equation}
for the $K'$ valley.
Thus the exchange of the valley won't change the quantum numbers $s_{z},\ s,\ n$.
$\Phi$ are the orthonormal eigenstates of the Harmonic oscillatior (one-dimension here) or the Landau level wave function,
and note that the $\Phi_{n}=0$ for $n<0$.
Also, the $(\Phi_{n}, \Phi_{n-1})^{T}$ is found to be the eigenstate of the operator 
$\frac{{\bf P}}{\sqrt{2}}(\sigma_{x}+i\sigma_{y})+\frac{{\bf P}^{\dag}}{\sqrt{2}}(\sigma_{x}-i\sigma_{y})$\cite{Hsu Y F}.

The resulting integer Hall conductivity is $\sigma_{xy}=2(2n+1)\frac{e^{2}}{h}$ for the case of zero Dirac-mass.
Here the parameter is $2(2n+1)=4n+1/2$, where the factor 4 here denotes the
number of degenerate (spin and valley) and $1/2$ is related to the pseudospin winding number,
which also implies the anomalous integer quantum Hall effect.
And it contains the filling factor $\nu=2n+1$,
which is different from that in the normal semiconductors which is $\nu=2n$.
What we focus on is the case of non-zero Dirac-mass,
which is corresponds to the Hall conductivity $\sigma_{xy}=(2n+1)\frac{e^{2}}{h}$, which is consistent with the result of Ref.\cite{Tahir M}
for zero disorder case.
However, this sequence of plateaus is easy to be affected by the disorder with a finite strength due to the dopping of impurities or the lattice defeats\cite{Liu Y L},
which we will discuss in the follwing section.
$E_{n}$ denotes the energy of the electron states (not the energy of hole state which is negative),
and it's
$E_{n}=\sqrt{2n\hbar^{2}v_{F}^{2}/\ell_{B}+(m^{\eta s_{z}}_{D})^{2}}$ for the $n\ne 0$ levels,
and $E_{0}=\eta m_{D}^{\eta s_{z}}$ for the zeroth-Landau level.

Under magnetic field, the width of the Landau level (the broaden of the energy level) as well as the quantized cyclotron orbit radius 
is proportional to disorder due to the exist of scattering rate $\Gamma=-{\rm Im}\Sigma(\omega)$,
whose sign is the same as that of the Fermi frequency $\omega$.
where the self-energy can be obtained self-consistently as\cite{Zimmermann R}
\begin{equation} 
\begin{aligned}
\Sigma(\omega)=-\sum_{{\bf q}}g({\bf q})[f({\bf k}+{\bf q})+\int^{\infty}_{-\infty}\frac{d\omega}{\pi}{\rm Im}\epsilon({\bf q},\omega)\frac{1-f({\bf k}+{\bf q})}
{\omega+\mu-\epsilon({\bf k}+{\bf q})}],
\end{aligned}
\end{equation}
which contains both the contributions of exchange and the fluctuation dissipations.
Thus the scattering rate is related to the Fermi frequency during the electron propagation,
and it's estimated as $\Gamma=0.1\hbar\omega_{c}$\cite{Shakouri K} here and scale as $\sim\sqrt{B}$ for the weak magnetic field,
and we also neglect the dependence on the Landau level, i.e., the $\Gamma$ is independs on $n$.

Under magnetic field and low-temperature (elastic scattering is dominate), 
the longitudinal conductivity is only allowed for the intra-Landau level transition,
however, in the present of large band gap opend by the intrinsic SOC,
the interband transitions are also limited.
In this case, the scattering wave vector is much smaller than the screened wave vector ${\bf k}_{s}$,
 which is polarization-dependent as ${\bf k}_{s}=2\pi e^{2}\Pi({\bf q},\omega)/(\epsilon_{0}\epsilon)$
and proportional to the Thomas-Fermi wave vector.
Then the screended scattering due to the charged impurities is governed by a $\delta$-like function
and the Coulomb potential becomes ${\bf q}$-independent.
The ${\bf q}$ is zero only for the elastic backscattering in which case the scattering potential is close to a $\delta$-function
similar to the Lorentzian representation\cite{Tabert C J2}
and become ${\bf q}$-independent.
In this case,
the longitudinal conductivity is 
\begin{equation} 
\begin{aligned}
\sigma_{xx}=\frac{4\pi^{2} e^{2}\Gamma}{TS_{B}h\omega_{c}}\sum_{n,s_{z},\eta}{\bf F}_{ss'}E_{n}f(E_{n})(1-f(E_{n})\delta_{E_{n'},E_{n}}(1-{\rm cos}^{2}\theta),
\end{aligned}
\end{equation}
where $\Gamma=\pi n^{{\rm imp}}V^{2}_{0}/\hbar S_{B}$ with the coulomb potential $V_{0}=e^{2}/2\epsilon_{0}\epsilon$,
and the scttering rate $\Gamma$ here is consistent with the two-dimension eklectron gas (2DEG)\cite{Grimaldi C} under the weak magnetic field.
The Coulomb interaction (scattering) matrix element here can be obtained from the three-term recurrence relation of the Laguerre polynomials in Favard's theorem 
\begin{equation} 
\begin{aligned}
(n+1)L^{\alpha}_{n+1}(r_{B})-(2n+1+\alpha-r_{B})L_{n}^{\alpha}(r_{B})+(n+\alpha)L_{n-1}^{\alpha}(r_{B})=0,
\end{aligned}
\end{equation}
where $\alpha=|n-n'|=0$ here since only account the intra-Landau level transition, and thus
\begin{equation} 
\begin{aligned}
&{\bf F}_{ss'}({\bf k},({\bf k}+{\bf q}))=\frac{1}{2}\left[1+ss'(\frac{{\bf k}({\bf k}+{\bf q})}{E_{{\bf k}}E_{{\bf k}+{\bf q}}}
+\frac{(m_{D}^{\eta s_{z}})^{2}}{E_{{\bf k}}E_{{\bf k}+{\bf q}}})\right]\\
&=\frac{1}{4}\left[(2n+1)(1+\frac{m_{D}^{\eta s_{z}}}{E_{n}})^{4}-2n(1+\frac{m_{D}^{\eta s_{z}}}{E_{n}})^{2}(1-\frac{m_{D}^{\eta s_{z}}}{E_{n}})^{2}
+(2n-1)(1-\frac{m_{D}^{\eta s_{z}}}{E_{n}})^{4}\right].
\end{aligned}
\end{equation}
While the above $\theta$ is the angle between ${\bf k}$ and ${\bf k}+{\bf q}$, it has
${\rm cos}\theta=\langle\chi({\bf k})|\chi({\bf k}+{\bf q})\rangle=(|{\bf k}|+|{\bf q}|{\rm cos}\phi)/|{\bf k}+{\bf q}|$ 
where $\phi$ is the angle between ${\bf k}$ and ${\bf q}$, and $|\chi({\bf k})\rangle=\psi_{s}^{*}({\bf k})\psi_{s'}({\bf k})$,
$|\chi({\bf k}+{\bf q})\rangle=\psi_{s}({\bf k}+{\bf q})\psi_{s'}^{*}({\bf k}+{\bf q})$ is the eigenstates with the eigenvectors $\psi$ of the Hamiltonian.
Note that here the scalar product are the simplification of $\langle\chi({\bf k})|\int^{\pi/a}_{-\pi/a}e^{-i{\bf q}r{\rm cos}\theta}d\theta|\chi({\bf k}')\rangle=
\langle\chi({\bf k})|\chi({\bf k}')\rangle \delta({\bf k}',{\bf k}+{\bf q})$.
The two-dimension impurities scattering potential after the Fourier transformation is $V({\bf k}_{s})=\frac{2\pi U}{\sqrt{({\bf q})^{2}+{\bf k}_{s}^{2}}}$
with the scattering wave vector $ {\bf q}$ and screening wave vector ${\bf k}_{s}$.
with the coulomb potential $U=\frac{qq'}{4\pi\epsilon_{0}\epsilon}$ where $\epsilon_{0}=1$ is the vacuum dielectric constant
and $\epsilon=2.5$ (air/SiO$_{2}$ substrate) is the background dielectric constant.
The scattering angle has ${\bf q}=|{\bf k}-{\bf k}'|=2k\ {\rm sin}\theta$\cite{Adam S,Vargiamidis V}, where $\theta$
describes the difference between the monentums before scattering and after scattering,
and it tends to zero $\theta\rightarrow 0$ for the SC silicene (deposited on a SC electrode or generate the topological superconductor by the STM probe).
The $\Delta {\bf k}$ is zero only for the elastic backscattering in which case the scattering potential is close to a $\delta$-function
similar to the Lorentzian representation
and become $\Delta {\bf k}$-independent.
In this case, the scattering potential is decay as $1/|{\bf k}_{s}|$.
Due to the exist of the impurities and lattice defects, the quantum spin-Hall effect with the spin-polarized current 
may be more observable due to the SOC with the impurities,
even without applying the external exchange field or the electric field,
and it's robust against the nonmagnetic impurity scattering.
The longitudinal conductivity here is purely real unlike
the expression of conductivity in Eq.(2) and Eq.(4)
which is complex as long as the scattering rate $\Gamma$ is variational for the different quantum number,
and that's also distincted from the magneto-optical conductivity or dynamical optical conductivity in 
silicene or graphene\cite{Tabert C J,Tabert C J2},
but it's correct for analytical evaluation here for low-temperature\cite{Krstajić P M}.

For the large ${\bf q}$ case and with the non-negligible effect from the magnetic field, the above scattering matrix element can be evalued as
\begin{equation} 
\begin{aligned}
&{\bf F}^{*}_{ss'}
=\int^{\infty}_{0}r_{B}e^{-r_{B}}\frac{[(1+\frac{m_{D}^{\eta s_{z}}}{E_{n}})^{2}L^{1}_{n}(r_{B})+(1-\frac{m_{D}^{\eta s_{z}}}{E_{n}})^{2}L^{1}_{n-1}(r_{B})]^{2}}{4}
dr_{B},
\end{aligned}
\end{equation}
for the $n\neq 0$ Landau levels,
and 
\begin{equation} 
\begin{aligned}
{\bf F}^{*}_{ss'}
=\int^{\infty}_{0}r_{B}e^{-r_{B}}dr_{B}
\end{aligned}
\end{equation}
for the zeroth Landau level.
The scattering rate now becomes 
\begin{equation} 
\begin{aligned}
\Gamma_{B}=\frac{U_{0}^{2}n^{{\rm imp}}}{4\pi r_{B}}=\frac{U_{0}^{2}n^{{\rm imp}}}{\pi \ell^{2}_{B}}
\end{aligned}
\end{equation}
in the full self-consistent Born approximation,
where the scattering momentum has ${\bf q}=1/\sqrt{2}$ here and since we put the magnetic field perpendicular to the silicene,
the $x$-component of scattering vector can be estimated as ${\bf q}_{x}=10^{6} m^{-1}$ in the experiment\cite{Krstajić P M},
thus the $\Gamma_{B}$ is proportional to $\sqrt{B}$.
In this case, the magnetic effects is much more important than the thermal spacing, $\hbar\omega_{c}\gg k_{B}T$,
and here $\hbar\omega_{c}$ is proportional to the electric field and the square root of the scattering rate $\Gamma_{B}$.
Then the resulting longitudinal conductivity can be obtained as
\begin{equation} 
\begin{aligned}
\sigma^{*}_{xx}=\frac{2\pi e^{2}\Gamma_{B}}{TS_{B}h^{2}\omega_{c}}\sum_{n,s_{z},\eta}{\bf F}^{*}_{ss'}E_{n}
[f(E_{n})(1-f(E_{n})+f(E_{n'})(1-f(E_{n'})]\delta_{E_{n'},E_{n}}(1-{\rm cos}^{2}\theta).
\end{aligned}
\end{equation}

At zero temperature, the spin and valley Hall conductivity has the regular form:
For the case of $\mu< m_{D}$ and $T=0$,
the spin Hall conductivity and valley Hall conductivity are\cite{Tahir M2}
\begin{equation} 
\begin{aligned}
\sigma_{xy}^{s}&=-\frac{e^{2}}{2h}[{\rm sgn}(\lambda_{SOC}+M-\frac{\overline{\Delta}}{2}E_{\perp})+{\rm sgn}(\lambda_{SOC}+M+\frac{\overline{\Delta}}{2}E_{\perp})],\\
\sigma_{xy}^{v}&=\frac{e^{2}}{2h}[{\rm sgn}(\lambda_{SOC}+M+\frac{\overline{\Delta}}{2}E_{\perp})-{\rm sgn}(\lambda_{SOC}+M-\frac{\overline{\Delta}}{2}E_{\perp})];
\end{aligned}
\end{equation}
while for the case of $\mu< m_{D}$ and $T=0$, they are
\begin{equation} 
\begin{aligned}
\sigma_{xy}^{s}&=-\frac{e^{2}}{2h}[\frac{\lambda_{SOC}+M-\frac{\overline{\Delta}}{2}E_{\perp}}{\sqrt{2n\hbar^{2}v_{F}/\ell_{B}+(\lambda_{SOC}+M-\frac{\overline{\Delta}}{2}E_{\perp})^{2}}}+
\frac{\lambda_{SOC}+M+\frac{\overline{\Delta}}{2}E_{\perp}}{\sqrt{2n\hbar^{2}v_{F}/\ell_{B}+(\lambda_{SOC}+M+\frac{\overline{\Delta}}{2}E_{\perp})^{2}}}],\\
\sigma_{xy}^{y}&=\frac{e^{2}}{2h}[\frac{\lambda_{SOC}+M+\frac{\overline{\Delta}}{2}E_{\perp}}{\sqrt{2n\hbar^{2}v_{F}/\ell_{B}+(\lambda_{SOC}+M+\frac{\overline{\Delta}}{2}E_{\perp})^{2}}}-
\frac{\lambda_{SOC}+M-\frac{\overline{\Delta}}{2}E_{\perp}}{\sqrt{2n\hbar^{2}v_{F}/\ell_{B}+(\lambda_{SOC}+M-\frac{\overline{\Delta}}{2}E_{\perp})^{2}}}],\\
\end{aligned}
\end{equation}
While for low-temperature (not zero),
the Hall conductivity of valley $K$ reads
\begin{equation} 
\begin{aligned}
&\sigma_{xy}^{K}=\frac{e^{2}}{h}\left\{\sum_{n}(n+\frac{1}{2})[f(E_{n})+f(-E_{n})-f(E_{n+1}-f(-E_{n+1}))]\right.\\
&\left.+\frac{1}{2}\sum_{n,\eta,s_{z}}m_{D}^{\eta s_{z}}(\frac{f(E_{n})-f(-E_{n}))}{E_{n}}-\frac{f(E_{n+1})-f(-E_{n+1}))}{E_{n+1}})\right\},
\end{aligned}
\end{equation}
and for valley $K'$, it reads
\begin{equation} 
\begin{aligned}
&\sigma_{xy}^{K}=\frac{e^{2}}{h}\left\{\sum_{n}(n+\frac{1}{2})[f(E_{n+1})+f(-E_{n+1})-f(E_{n}-f(-E_{n}))]\right.\\
&\left.+\frac{1}{2}\sum_{n,\eta,s_{z}}m_{D}^{\eta s_{z}}(\frac{f(E_{n+1})-f(-E_{n+1}))}{E_{n+1}}-\frac{f(E_{n})-f(-E_{n}))}{E_{n}})\right\},
\end{aligned}
\end{equation}
For zeroth Landau level, which is thought to be in the bottom of the conduction band and with zero energy here,
has only half of the degeneracy of the other levels and depending on the sign of $e{\bf B}$.
It's
not in the same place for valleys $K$ and $K'$ except for the particle-hole symmetry case,
and it's either occupaied by the electron or the hole,
however, to understanding the anomalous integer quantum Hall effect compared to the normal semiconductors,
the assumption of a zero energy-level which shared by both the conduction band and valence band\cite{Shakouri K} is useful. 
The gap does not affects the longitudinal conductivity in the zeroth-Landau level,
since they cancer each other in such case\cite{Shakouri K}.
Here we note that, our above results can also be applied to the bilayer silicene, which
just need to replace the above eigenvalue ($E_{n}$) by the 
$E_{n}=\sqrt{2n\hbar^{2}v_{F}^{2}/\ell_{B}+(m^{\eta s_{z}}_{D})^{2}}\pm t^{{\rm inter}}$,
where sign $\pm$ stands the upper layer and lower layer, respectively,
where the $t^{{\rm inter}}$ is the interlayer hopping which we has been investigated in the previous works\cite{Wu C H,Wu C H2,Wu C H3}.
and the perpendicular conductivity\cite{Tabert C J2} can be obatined throught the $z$-direction velocity operator
$v_{z}=t^{{\rm inter}}d\sigma_{z}/\hbar$ with the interlayer distance $d$ and the $z$-component Pauli operator.

Finally, we know that the density of states in the silicene is related to the value of dynamical conductivity and the dynamical polarization
esperially in the long-wavelength limit (small ${\bf q}$ limit).
Then we deduce the dynamical polarization under different conditions.
In static case, in which both the conductivity and polarization doesn't depends on the fermionic Matsubara frequency,
\begin{equation} 
\begin{aligned}
\Pi({\bf q},0)=-g_{s}g_{v}\frac{2e^{2}\mu}{2\pi\epsilon_{0}\epsilon \hbar^{2} v_{F}^{2}}
\left[\frac{ m_{D}}{2\mu}+\frac{\hbar^{2} v_{F}^{2}{\bf q}^{2}-4 (m_{D}^{\eta s_{z}})^{2}}{4\hbar v_{F}{\bf q}\mu}{\rm arcsin}\sqrt{\frac{\hbar^{2}v_{F}^{2}{\bf q}^{2}}{\hbar^{2}v_{F}^{2}{\bf q}^{2}+4 m_{D}^{2}}}\right]
\end{aligned}
\end{equation}
for $0<\mu< m_{D}^{\eta s_{z}}$,
\begin{equation} 
\begin{aligned}
\Pi({\bf q},0)=-g_{s}g_{v}\frac{2e^{2}\mu}{2\pi\epsilon_{0}\epsilon \hbar^{2} v_{F}^{2}}
\left[1-\Theta({\bf q}-2{\bf k}_{F})\left(\frac{\hbar^{2} v_{F}^{2}\sqrt{{\bf q}^{2}-4{\bf k}_{F}^{2}}}{2\hbar v_{F}{\bf q}}-\frac{\hbar^{2}v_{F}^{2}{\bf q}^{2}-4 m_{D}^{2}}{4\mu \hbar v_{F}{\bf q}}{\rm arctan}\frac{\hbar v_{F}\sqrt{{\bf q}^{2}-4{\bf k}^{2}_{F}}}{2\mu}\right)\right]
\end{aligned}
\end{equation}
for $\mu> m_{D}^{\eta s_{z}}$.
While in the absence of Dirac-quasiparticle scattering and with nonzero temperature\cite{Gorbar E V},
the static polarization reads
\begin{equation} 
\begin{aligned}
\Pi(0,0)=-g_{s}g_{v}\frac{2e^{2}T}{2\pi\epsilon_{0}\epsilon v_{F}^{2}}\left[{\rm ln}(2{\rm cosh}\frac{m_{D}+\mu}{T})-\frac{m_{D}}{2T}{\rm tanh}\frac{m_{D}+\mu}{2T}+(\mu\rightarrow -\mu)\right].
\end{aligned}
\end{equation}
In the zero-temperature limit, it becomes proportional to the density of states $D$ of the Dirac-quasiparticle which is a step function now
\begin{equation} 
\begin{aligned}
\Pi(0,0)_{T\rightarrow 0}=
-e^{2}D(|\mu|)=-e^{2}\frac{g_{s}g_{v}|\mu|}{2\pi\hbar^{2}v_{F}^{2}}\frac{1}{2}\sum_{\eta=\pm 1}\left[\theta(|2\mu|-2|m_{D}|_{\eta})\right].
\end{aligned}
\end{equation}
Since we assume the scattering rate is independs on the Landau level index $n$ due to the large ${\bf q}$ character,
that, at large momentum, the effect of magnetic-field is been weakened,
and the Landau level index can then even be replaced by the $r_{B}$ in the zero scattering case\cite{Pyatkovskiy P K}.
While in the long-wavelength case (small ${\bf q}$), the transitions between different Landau level index $n$ is important 
for the scattering rate.
The scattering rate what we use here is still $\Gamma_{B}$.
The resulting dynamical polarization in one-loop approximation and in non-static case with finite scattering wave vector is\cite{Pyatkovskiy P K}
\begin{equation} 
\begin{aligned}
\Pi({\bf q},\omega)=-g_{s}g_{v}\frac{e^{2}}{4\pi\ell_{B}^{2}}\sum_{n.n',s,s'}\frac{f(sE_{n})-f(s'E_{n'})}{sE_{n}-s'E_{n'}+\omega+i\delta+i\Gamma}Q(m_{D}^{\eta,s_{z}})
\end{aligned}
\end{equation}
where we ignore the $n$-dependence of the scattering rate and set the frequency as $1$ here,
the Dirac-mass term $Q(m_{D}^{\eta,s_{z}})$ is
\begin{equation} 
\begin{aligned}
Q(m_{D}^{\eta,s_{z}})=&e^{-r_{B}}r_{B}\left[(1+\frac{ss'(m_{D}^{\eta,s_{z}})^{2}}{E_{n}E_{n'}})(\frac{n!}{n'!}(L^{1}_{n}(r_{B}))^{2}
+(1-\delta_{0n})\frac{(n-1)!}{(n'-1)!}(L^{1}_{n-1}(r_{B}))^{2})\right.\\
&\left.+\frac{4ss'v_{F}^{2}\hbar^{2}}{\ell_{B}E_{n}E_{n'}\frac{n!}{(n'-1)!}}L^{1}_{n-1}(r_{B})L^{1}_{n}(r_{B})\right].\\
\end{aligned}
\end{equation}

\section{Simulation results and discussions}

We first focus on the zero-gap ($m_{D}=0$) case where the electric field is 0.034 eV.
In Fig.1(a), we show the magneto-conductivity with electric field is 0.034 eV,
and under the magnetic field $B=1$ T, low temperature $T=0.1$ K.
The resulting longitudinal conductivity and the valley contribution are real,
only the Hall conductivity is complex.
Note that in Fig.1, for $\sigma_{xy}$, we set the scattering rate $\Gamma=0.1\hbar\omega_{c}$ 
(correspons to $n^{{\rm imp}}V^{2}({\bf q})=4.23\times 10^{9}$),
while for $\sigma_{xx}$ we set the scattering rate as $\Gamma=n^{{\rm imp}}\pi U_{0}/\hbar$
which is similar with the 2DEG under the weak magnetic field.
The $\sigma_{xx}$ shown in the Fig.1 contains the intra-Landau level transition (accounts the $n$ level only)
due to the domination of elastic scattering,
further, since we applied exchange field in the small electric field case (Fig.1(a))
which is 0.034 eV now and leads to the gapless band struture,
there is a large peak in the zero chemical potential place (the charge-neutral point).
This large peak shows not oscillation behavior in the charge-neutral point,
and thus it's independs on the electron concentration,
unlike the other longitudinal conductivity or the resistivity\cite{Krstajić P M} which with finite $\mu$.
In fact, for gapless case as shows in this panel,
the $\sigma_{xx}$ won't vanish but exhibit a minimum plateau as $e^{2}/h$ (here is for the single spin component and valley component case; i.e.,
don't consider the spin and valley degenerate),
which can be seen in the real part of the Hall conductivity as shown in the Fig.1(a) and Fig.2(a).
This minimum plateau is independent of the Dirac-point singularity\cite{Krstajić P M},
but depends on the carrier imbalance and the applied magnetic field like the filling factor as shown in Ref.\cite{Gusynin V P}.

While for the larger electric field as 0.0064 eV (Fig.1(b)) the peak in charge-neutral point is vanish and the spin degeneracy is lifted,
the resulting band gap opened by the collective effect of intrinsic SOC, applied exchange field and the electric field,
exclude the interband transition (it will be more clear if the band gap is large enough),
and results in the invariance of the quantum number of $s_{z}$, $s$, and $\eta$ during the scattering,
(The Zeeman splitting was neglected here since it's negligible in quantity until the magnetic field is larger than 20 T).
In Fig.2, we present the result of $\sigma_{xy}$ with the impurities concentration
$n^{{\rm imp}}V^{2}({\bf q})=2.65\times 10^{10}$,
We surprised to find that the real part Hall conductivity ${\rm Re}[\sigma_{xy}]$ is close to the valley contribution $\sigma_{xy}^{K}$,
but the conductivity also damping more quickly due to the increasing of the impurities concentration also 

Fig.3 shows the longitudinal conductivity (labeled as $\sigma_{xx}^{*}$) with the inter-Landau level transitions,
which contains ont only the elastic scattering.
We found that in this case the longitudinal conductivity becomes double-peak,
which is due to the inter-Landau level transition,
even for the $n=0$ Landau level,
and the double peaks in the $n=0$ level are generally much larger that other levels.
The first peak of the $n=0$ level is in the charge-neutral point for gapless case and not in the gapped case,
which is the same as the $\sigma_{xx}$ in Fig.1 and Fig.2.
The double peak longitudinal conductivity shown in Fig.3 where the scattering rate is setted as 
$\Gamma=0.1\hbar\omega_{c}$ ($n^{{\rm imp}}V^{2}({\bf q})=4.23\times 10^{9}$),
corresponds to the inter-Landau level transition from $n$ to $n'=n+1$,
and its peak is proportional to the scattering rate.
That can be proven by the $B=2$ T curve, that when the magnetic field $B$ is increased to 2T, 
the magnetic-dependent parameter $r_{B}$ decreased and then enhance the scattering rate,
we can see the resulting $\sigma_{xx}$ is much larger than the $B=1$ one.
That can also be explainted by the increasing of the inter-Landau level transition,
and the result here is consistent with Refs.\cite{Krstajić P M,Shakouri K}.
We can also see that, when increasing the electric field to 5 eV, the longitudinal conductivity in $n=0$ level is heavily reduced,
while the $n\neq 0$ levels are changed slightly,
but becomes larger and shifted towards the right compared to the $E_{\perp}=0.034$ eV one.

Through Fig.4, we can see that the non-static $\Pi({\bf q},\omega)$ is complex while static long-wavelength one $\Pi(0,0)$ is real,
and the $\Pi({\bf q},\omega)$ is depends both on the electric field and the magnetic field, while $\Pi(0,0)$ only depends on the electric field.
From Fig.4(a), we found that
when the electric field increase, both the real part and the imaginary part of polarization shift towards the right.
We also found that,  under the magnetic field, the polarization function 
which is logarithmically divergen when under zero magnetic field (as shown in our previous work\cite{Wu C H3}),
becomes step-like and Landau level-dependent.
In $B=1$ T, there are two peaks for the real polarization, while there is only one peak for the imaginary part of the polarization.
In $B=2$ T, the two-peak-feature vanish, and the polarization function act more like the Hall conductivity,
i.e., the step-like feature is more obviously.
The static polarization function in long-wavelength limit (${\bf q}\rightarrow 0$)
and in the absence of Dirac-quasiparticle scattering with finite temperature,
is shown in the Fig.4(b), we found that it exhibits linear relation in the large chemical potential region
and the slope is independent of the electric field or magnetic field,
but only related to the temperature.

In this papaer we analytically investigate the conductivity of the silicene with a finite chemical potential,
the integer quantum Hall conductivity and longitudinal conductivity are explored.
We not only investigate the low-temperature and $\delta$-function impurities (i.e., independent of the scattering momentum) case,
which only exist the intra-Landau level transition,
but also for the case of inter-Landau level transition which also with the non-elastic scattering.
We also compute the dynamical polarization under the magnetic field 
which has many exciting and novel properties, 
and with the screened scattering due to the charged impurities.
In fact, the screening here may also due to the plasmon collective model which we have discussed in other place\cite{Wu C H}.
The polarization function is related to the interband and intraband transition properties,
but under magnetic field,
it becomes also related to the Landau level index 
and shows step-like feature but not the logarithmically divergenas, as shown above,
and it's also naturally related to the conductivity in silicene.
The generalized Laguerre polynomial is used in both the longitudinal conductivity and dynamical polarization function
under magnetic field.
For the longitudinal conductivity studied in this letter, we takes account the collision of carriers with large charged impurities,
and ignore the unimportant diffusive effect, and thus it's also related to the Shubnikov-de Haas oscillation.
In fact, for the bilayer silicene which with quadratic dispersion near the Dirac-point,
the electron-electron interaction doesn't affect the cyclotron energy under the magnetic field
in equilibium electron system due to the Kohn theorem\cite{Kohn W},
but for the monolayer silicene which with linear dispersion near the Dirac-point under the magnetic field,
the case is opposite.

\end{large}
\renewcommand\refname{References}

\clearpage
Fig.1
\begin{figure}[!ht]
   \centering
   \begin{center}
     \includegraphics*[width=0.8\linewidth]{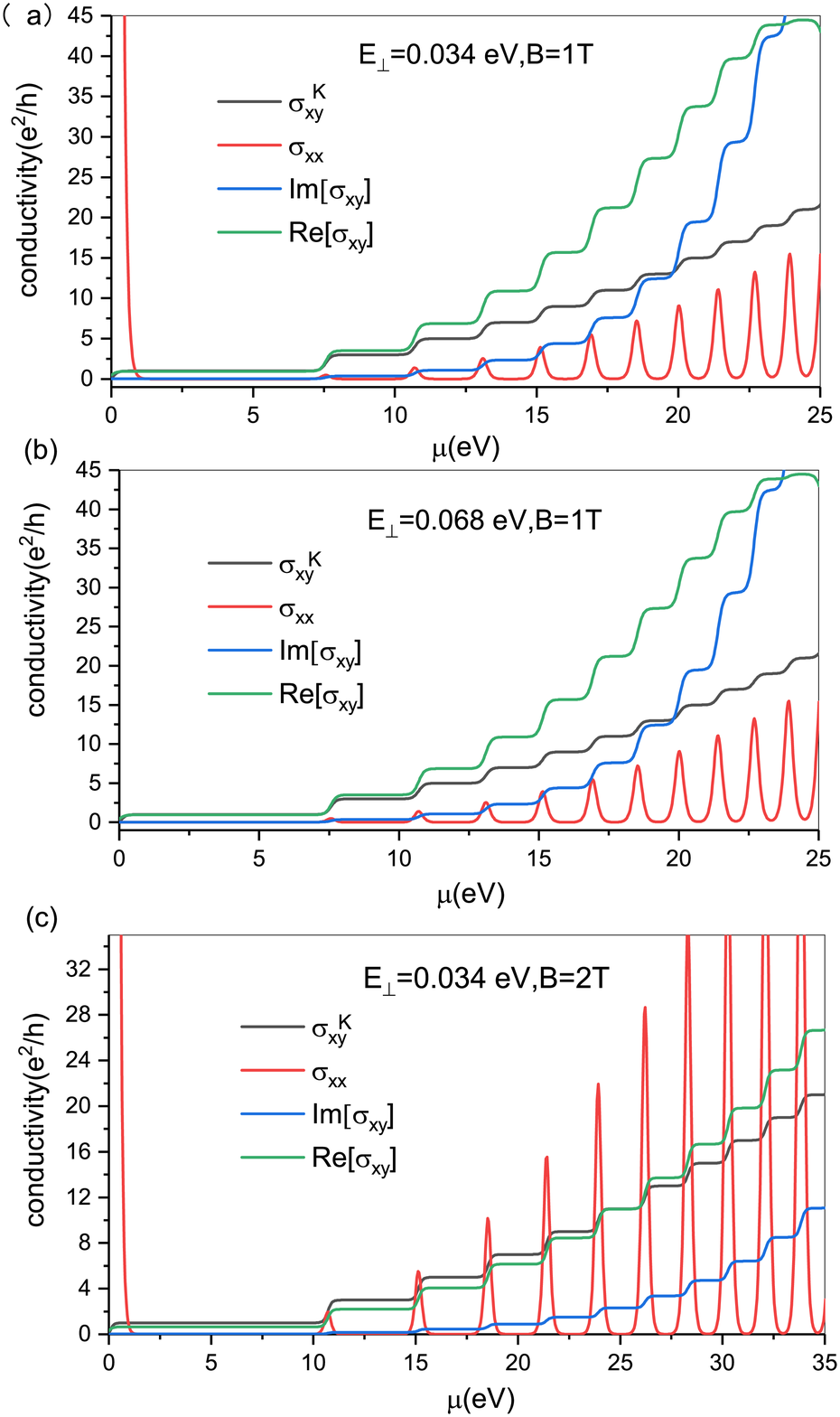}
\caption{(Color online) Real part and imaginary part of the Hall conductivity $\sigma_{xy}$, longitudinal conductivity $\sigma_{xy}$,
and valley contribution to the Hall conductivity $\sigma_{xy}^{K}$ as a function of the chemical potential (Fermi energy) $\mu$.
For $\sigma_{xy}$, we set the scattering rate $\Gamma=0.1\hbar\omega_{c}$ which correspond to
$n^{{\rm imp}}V^{2}({\bf q})=4.23\times 10^{9}$.
Here we set the $\hbar=1$, $T=0.1$ K, $B=1$ T.
For $sigma_{xx}$ (only consider the intra-Landau level transition here) the scattering rate is $\Gamma=\pi n^{{\rm imp}}V_{0}^{2}/\hbar S_{B}$.
The corresponding electric field and magnetic field are labeled in each panels.
}
   \end{center}
\end{figure}
\clearpage

Fig.2
\begin{figure}[!ht]
   \centering
   \begin{center}
     \includegraphics*[width=0.8\linewidth]{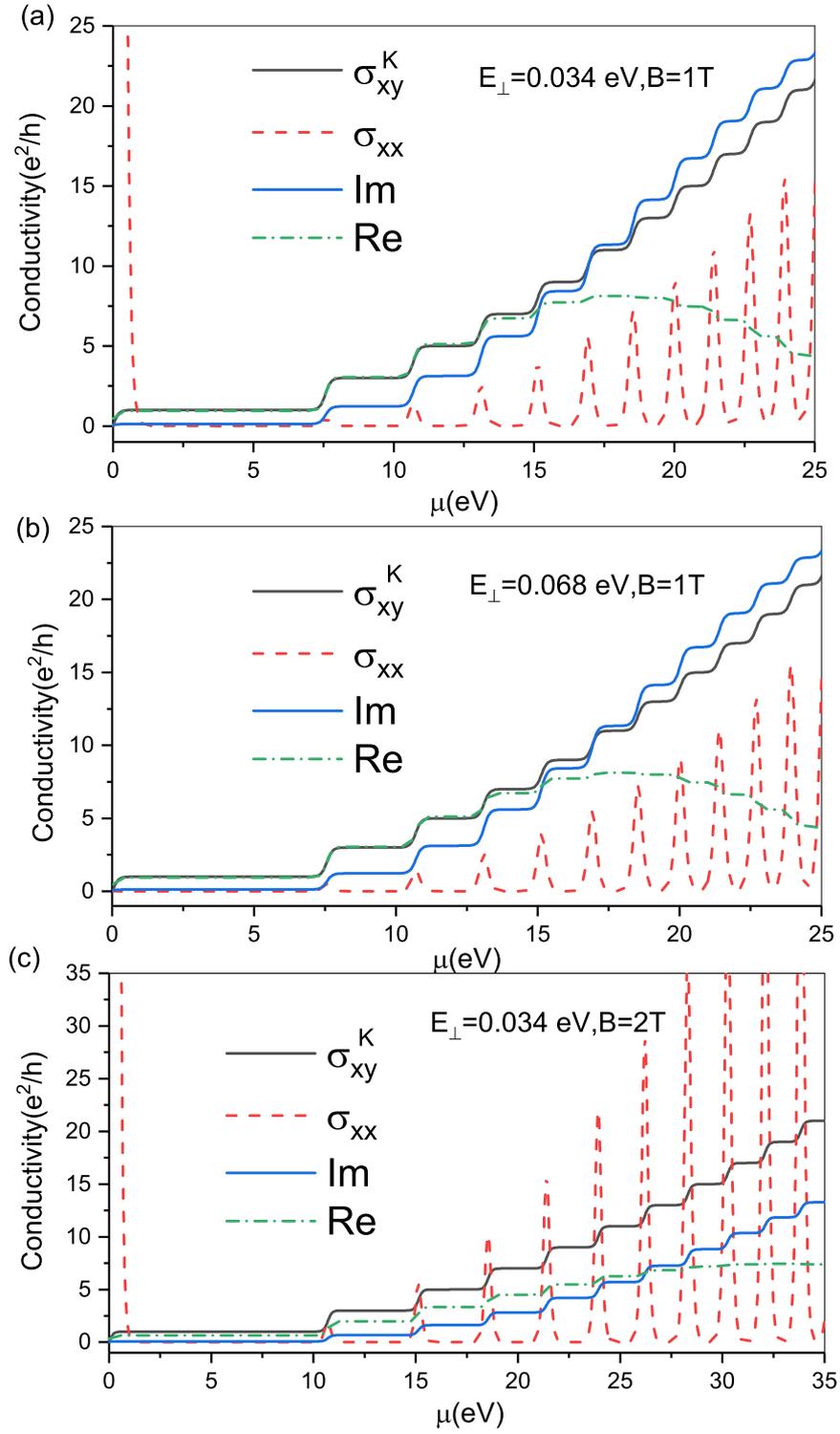}
\caption{(Color online) The same as in Fig.1 but for $n^{{\rm imp}}V^{2}({\bf q})=2.65\times 10^{10}$.
}
   \end{center}
\end{figure}
\clearpage
Fig.3
\begin{figure}[!ht]
   \centering
   \begin{center}
     \includegraphics*[width=0.8\linewidth]{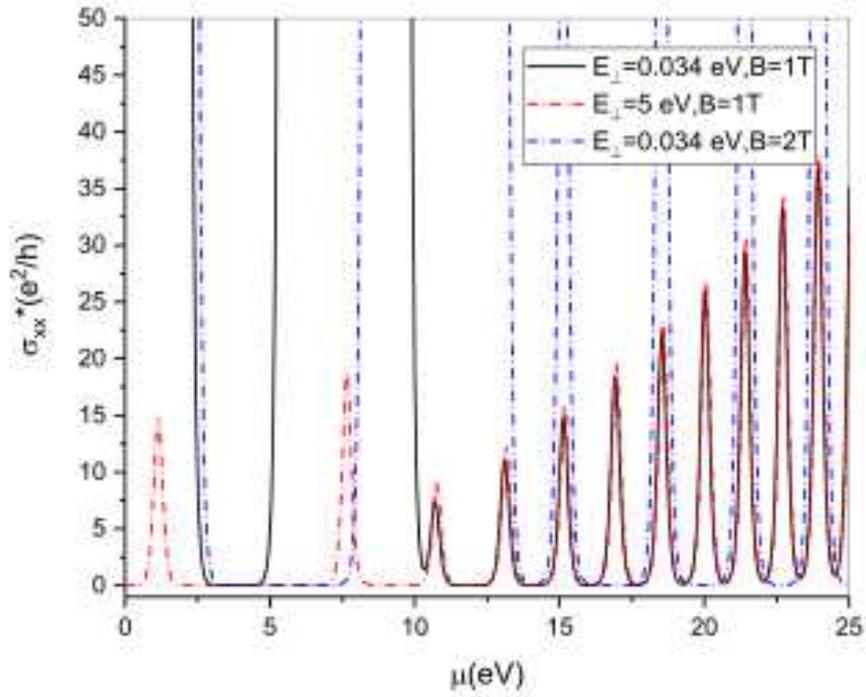}
\caption{(Color online) The longitudinal conductivity $\sigma_{xx}^{*}$ under the electric field and magnetic field as a function of the chemical potential,
which contains the inter-Landau level transition and non-elastic scattering here.
}
   \end{center}
\end{figure}
\clearpage
Fig.4
\begin{figure}[!ht]
   \centering
   \begin{center}
     \includegraphics*[width=0.8\linewidth]{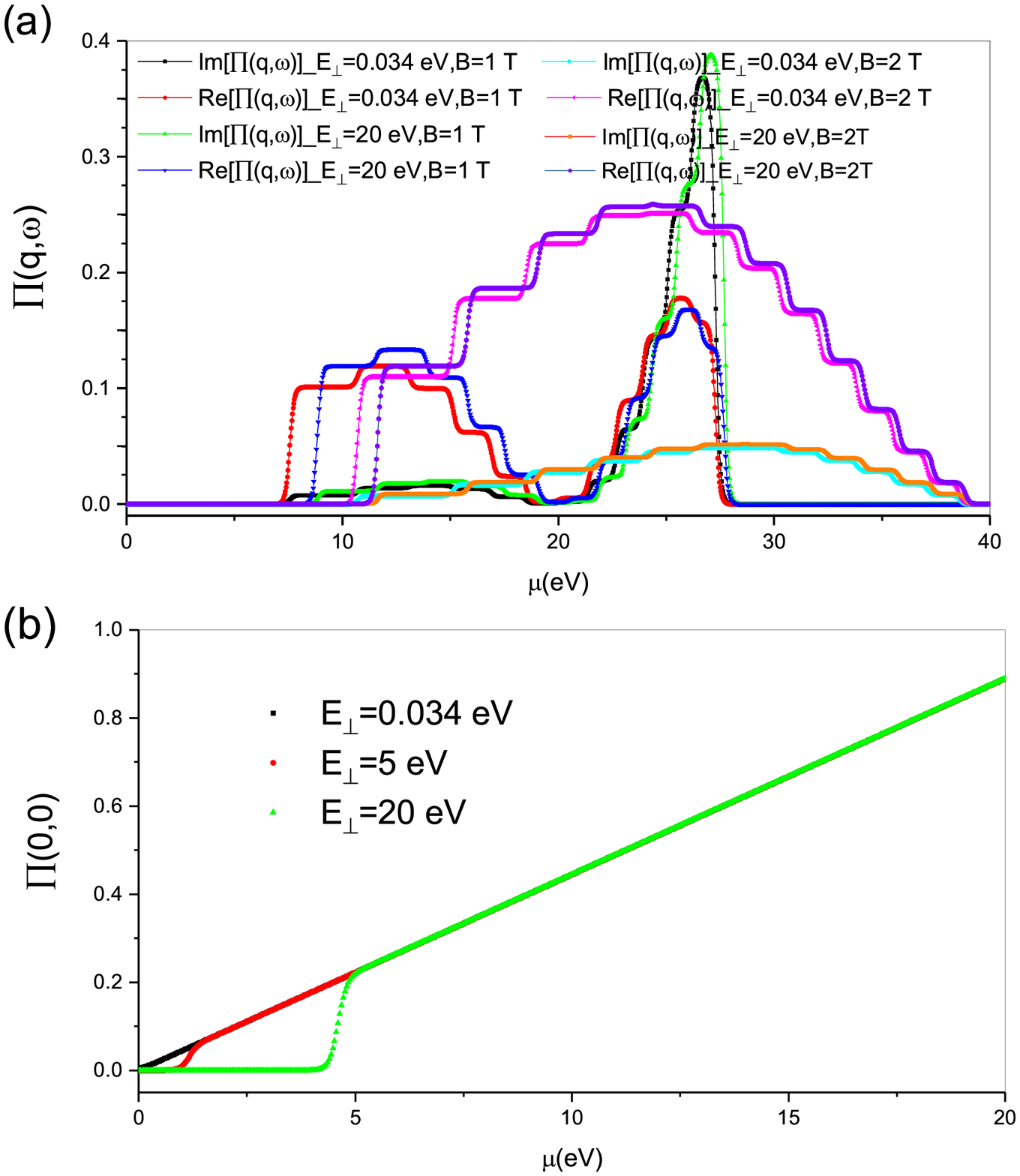}
\caption{(Color online) Non-static polarization function $\Pi({\bf q},\omega)$ (a) and the static one in long-wavelength case $\Pi(0,0)$ (b)
under different electric field and magnetic field.
The temperature is setted 1 T here and the Fermi frequency setted as 1.
}
   \end{center}
\end{figure}

\begin{thebibliography}{99}
\bibitem{Feng B}Feng B, Ding Z, Meng S, et al. Evidence of silicene in honeycomb structures of silicon on Ag (111)[J]. Nano letters, 2012, 12(7): 3507-3511.
\bibitem{Ezawa M}Ezawa M. Valley-polarized metals and quantum anomalous Hall effect in silicene[J]. Physical review letters, 2012, 109(5): 055502.
\bibitem{Tao L}Tao L, Cinquanta E, Chiappe D, et al. Silicene field-effect transistors operating at room temperature[J]. Nature nanotechnology, 2015, 10(3): 227.
\bibitem{Guinea F}Guinea F. Spin-orbit coupling in a graphene bilayer and in graphite[J]. New Journal of Physics, 2010, 12(8): 083063.
\bibitem{Charbonneau M}Charbonneau M, Van Vliet K M, Vasilopoulos P. Linear response theory revisited III: One‐body response formulas and generalized Boltzmann equations[J]. Journal of Mathematical Physics, 1982, 23(2): 318-336.
\bibitem{Kotov V N}Kotov V N, Uchoa B, Pereira V M, et al. Electron-electron interactions in graphene: Current status and perspectives[J]. Reviews of Modern Physics, 2012, 84(3): 1067.
\bibitem{Wu C H}Wu C H. Geometrical structure and the electron transport properties of monolayer and bilayer silicene near the semimetal-insulator transition point in tight-binding model[J]. arXiv preprint arXiv:1805.00350, 2018. 
\bibitem{Tabert C J}Tabert C J, Nicol E J. Magneto-optical conductivity of silicene and other buckled honeycomb lattices[J]. Physical Review B, 2013, 88(8): 085434.
\bibitem{Wu C H2}Wu C H. Tight-binding model and ab initio calculation of silicene with strong spin-orbit coupling in low-energy limit[J]. arXiv preprint arXiv:1804.01695, 2018.
\bibitem{Hsu Y F}Hsu Y F, Guo G Y. Anomalous integer quantum Hall effect in A A-stacked bilayer graphene[J]. Physical Review B, 2010, 82(16): 165404.
\bibitem{Tahir M}Tahir M, Schwingenschlögl U. Valley polarized quantum Hall effect and topological insulator phase transitions in silicene[J]. Scientific reports, 2013, 3: 1075.
\bibitem{Liu Y L}Liu Y L, Luo G X, Xu N, et al. Integer quantum Hall effect and topological phase transitions in silicene[J]. arXiv preprint arXiv:1712.05348, 2017.
\bibitem{Zimmermann R}Zimmermann R, Kilimann K, Kraeft W D, et al. Dynamical screening and self‐energy of excitons in the electron–hole plasma[J]. physica status solidi (b), 1978, 90(1): 175-187.
\bibitem{Shakouri K}Shakouri K, Vasilopoulos P, Vargiamidis V, et al. Integer and half-integer quantum Hall effect in silicene: Influence of an external electric field and impurities[J]. Physical Review B, 2014, 90(23): 235423.
\bibitem{Tabert C J2}Tabert C J, Nicol E J. Dynamical conductivity of AA-stacked bilayer graphene[J]. Physical Review B, 2012, 86(7): 075439.
\bibitem{Grimaldi C}Grimaldi C, Cappelluti E, Marsiglio F. Off-Fermi surface cancellation effects in spin-Hall conductivity of a two-dimensional Rashba electron gas[J]. Physical Review B, 2006, 73(8): 081303.
\bibitem{Adam S}Adam S, Hwang E H, Galitski V M, et al. A self-consistent theory for graphene transport[J]. Proceedings of the National Academy of Sciences, 2007, 104(47): 18392-18397.
\bibitem{Vargiamidis V}Vargiamidis V, Vasilopoulos P, Hai G Q. Dc and ac transport in silicene[J]. Journal of Physics: Condensed Matter, 2014, 26(34): 345303.
\bibitem{Krstajić P M}Krstajić P M, Vasilopoulos P. Integer quantum Hall effect in gapped single-layer graphene[J]. Physical Review B, 2012, 86(11): 115432.
\bibitem{Tahir M2}Tahir M, Manchon A, Sabeeh K, et al. Quantum spin/valley Hall effect and topological insulator phase transitions in silicene[J]. Applied Physics Letters, 2013, 102(16): 162412.
\bibitem{Wu C H3}Wu C H. Interband and intraband transition, dynamical polarization and screening of the monolayer and bilayer silicene in low-energy tight-binding model[J]. arXiv preprint arXiv:1805.07736, 2018. 
\bibitem{Gorbar E V}Gorbar E V, Gusynin V P, Miransky V A, et al. Magnetic field driven metal-insulator phase transition in planar systems[J]. Physical Review B, 2002, 66(4): 045108.
\bibitem{Dyakonov M I}Dyakonov M I, Perel V I. Current-induced spin orientation of electrons in semiconductors[J]. Physics Letters A, 1971, 35(6): 459-460.
\bibitem{Krstajić P M}Krstajić P M, Vasilopoulos P. Integral quantum Hall effect in graphene: Zero and finite Hall field[J]. Physical Review B, 2011, 83(7): 075427.
\bibitem{Pyatkovskiy P K}Pyatkovskiy P K, Gusynin V P. Dynamical polarization of graphene in a magnetic field[J]. Physical Review B, 2011, 83(7): 075422.
\bibitem{Gusynin V P}Gusynin V P, Sharapov S G. Unconventional integer quantum Hall effect in graphene[J]. Physical review letters, 2005, 95(14): 146801.
\bibitem{Kohn W}Kohn W. Cyclotron resonance and de Haas-van Alphen oscillations of an interacting electron gas[J]. Physical Review, 1961, 123(4): 1242.

\end{thebibliography}
\end{document}